\documentclass[%
 reprint,
superscriptaddress,
 amsmath,amssymb,
 aps,
 longbibliography
]{revtex4-1}

\usepackage{mathtools}

\DeclarePairedDelimiter\floor{\lfloor}{\rfloor}

\usepackage[usenames]{color}
\usepackage{amssymb}
\usepackage{amsmath} 

\usepackage{esdiff}
\usepackage{amsthm}
\usepackage{placeins}
\usepackage{makeidx}
\usepackage{url}
\usepackage{float}
\usepackage[caption = false]{subfig}
\usepackage[normalem]{ulem}
\usepackage{tabto}

\usepackage{graphicx}
\usepackage{dcolumn}
\usepackage{bm}

\usepackage[mathlines]{lineno}

\usepackage{etoolbox}
\makeatletter
\patchcmd\linenumberpar{\@LN@parpgbrk}{\penalty\@LN@parpgpen\relax}{}{}
\makeatother

\begin{document}


\title{
Effects of memory on spreading processes in non-Markovian temporal networks
}

\author{Oliver E. Williams}
\affiliation{School of Mathematical Sciences, Queen Mary University of London, London, E1 4NS, United Kingdom}

\author{Fabrizio Lillo}
\affiliation{Department of Mathematics, University of Bologna, Piazza di Porta San Donato 5, 40126, Bologna, Italy}

\author{Vito Latora}
\affiliation{School of Mathematical Sciences, Queen Mary University of London, London, E1 4NS, United Kingdom}

\affiliation{Dipartimento di Fisica ed Astronomia, Universit\`a di Catania and INFN, I-95123 Catania, Italy}

\affiliation{Complexity Science Hub Vienna (CSHV), Vienna, Austria}


\begin{abstract}

  Many biological, social and man-made systems are better described in
  terms of temporal networks, i.e.~networks whose links are only
  present at certain points in time, rather than by static ones. In
  particular, it has been found that non-Markovianity is a necessary
  ingredient to capture the non-trivial temporal patterns of
  real-world networks. However, our understanding of how memory can
  affect the properties of dynamical processes taking place over
  temporal networks is still very limited, being especially constrained to 
  the case of short-term memory. Here, by introducing a
  model for temporal networks in which we can precisely control the
  link density and the strength and length of memory for each link, we
  unveil the role played by memory on the dynamics of epidemic
  spreading processes. Surprisingly, we find that the average spreading
  time in our temporal networks is often non-monotonically dependent on the
  length of the memory, and that the optimal value of the memory length
  which maximizes the spreading time depends on the strength of the
  memory and on the density of links in the network.  
  Through analytical arguments we then
  explore the effect that changing the number and length of network
  paths connecting any two nodes has on the value of optimal memory.
\end{abstract}

\date{\today}

\maketitle

\section{INTRODUCTION}

When a system is composed of many individual entities and pairwise 
interactions between them, then it is natural to describe its underlying structure 
as a complex network. We then say that it is on this backbone that 
all relevant dynamical processes take place \cite{Newman10,Latora_book17}. 
Often in real world systems this underlying structure is 
in its self dynamic, and so it is better
described in terms of networks in which links among a fixed
set of nodes change over time \cite{Holme_rev12,Holme_Temp_net_book,masuda_guide_temp_net,Mazzarisi08}.
Examples of such temporal networks include human contacts, which vary as individuals move over space
\cite{Gonzalez08,Buscarino08}, online social interactions that take 
place at certain points in time \cite{Holme05}, or functional brain networks where
correlations among the different areas of the human brain fluctuate
over time \cite{Valencia08,Fallani08}.
Recently many of these, and similar, systems have been empirically investigated,  
and the main dynamical properties required of temporal 
networks in order for them to better describe reality 
have started to be uncovered \cite{Tang:2010}.
In particular, it has been found that
non-Markovianity is necessary to capture the non-trivial
temporal patterns of real-world networks 
\cite{Kim_epj15,Rosvall_natcomm14,Scholtes_natcomm14,Scholtes_sigkdd17},
and can play an important role on processes occurring on temporal networks. 
It has been shown that non-Markovianity  
can affect the dynamics of random walks
\cite{Lambiotte_jcn15}, the speed of information 
\cite{Scholtes_natcomm14}, and the way diseases 
spread in systems with non-exponential inter-event
times \cite{Van_Mieghem_PRL13,Kiss_PRL15,Guening_EPJ15,Starnini_PRL17,Onaga_PRL17,kiss_book17}.
Also non-Markovianity turns out to be useful in the definition
of flow-based communities \cite{Salnikov_scirep16}.

While the presence of memory has been found to be important, 
the influence that the strength (intensity) and length (range or order)
of this memory  
can have on dynamical processes are still poorly understood.
To shed light on the role of memory on spreading processes on
temporal networks, we here propose a model for 
time-varying networks where the
dynamics of the links is non-Markovian and is generated by a discrete
autoregressive process of tuneable order. The key feature of our
model, which we call the \emph{Discrete Autoregressive Network}  
model of order $p$, or DARN(p) model,
is that it allows us to precisely control not only the graph density, but also
the strength and length of memory of each link.
By considering a standard 
susceptible-infected (SI) epidemic over networks generated by the proposed model, we study how the range of the memory affects the rate at which infection
is spread across the network. The main result of our work is to show that memory can play either of two opposing
roles: it can slow-down or speed-up the spreading depending on the features
of the network. 
In particular, it turns out that that the average spreading
time is often non-monotonically dependent on
the memory length, and there is a given value of the memory
length for which we obtain the maximal average time until the entire network is
infected. The DARN(p) model as presented is 
analytically tractable enough to allow for a more in depth study 
of the influence of memory than would be practical through 
numerics alone. We are in fact able to predict through analytical
arguments the position of this maximum for a range of values 
of the variables associated with the model.
We then explore the 
effect that changing the number and length of network paths 
connecting two nodes has on the value for the memory 
length which maximises the average passage time of the infection.

\section{MODELLING TEMPORAL NETWORKS WITH MEMORY}
Generating temporal networks is conceptually simple and a great deal
of work has been done in this area
\cite{Grindrod_prslA10,Holme_rev12}. One takes a set of nodes and
defines some way for them to interact over time (be it discrete or
continuous). Without the direct use of empirical data, this can be
done in a number of different ways, each with their own advantages and
disadvantages. For instance, much attention has been devoted to
temporal networks generated by the interactions of individuals in
agent based models
\cite{Buscarino:2008,Tang:2010,Starnini:2013,Karsai:2014aa,Perra:2012ab}.
While such models might intuitively reflect reality on some level,
they are often difficult to work with, without relying entirely on
simulations. 
In order to keep precise control over key aspects of a temporal
network, such as the strength and length of memory, while maintaining
analytical tractability, we propose a model for generating temporal
networks by assigning each link its own independent stochastic process.
Given a set of nodes
$\cal N$, with $|{\cal N}|=N$, we assign to each possible pair $i,j
\in {\cal N}$ a discrete time stochastic process for the element of
the adjacency matrix $X_t^{ij}$ such that $X_t^{ij} \in \{0,1\}
~\forall t$. For our purposes, we take links to be undirected, and
for any two different links, $(i,j)$ and $(k,l)$, the two random variables 
$X_t^{ij}$ and $X_t^{kl}$ are independent and identically distributed (i.i.d). 
Thus we can talk more generally about the single process $X_t$ without
worrying about which link we are referring to. The most important
ingredient of our model is that the process $X_t$ 
is in general not only non-Markovian, but has a precisely 
controllable amount of memory. This in practice means that 
the presence of link $(i,j)$ at time $t_2$ can depend on the
presence of the link at time $t_1$ for any $t_2 > t_1$. 
In particular, $X_t$ is 
chosen as a special case of the Discrete Auto-Regressive Process of
fixed order $p$, from now on referred to as DAR($p$) 
\cite{jacobs1978discrete,macdonald1997hidden,Mazzarisi08}, which allows us to control
both for the strength of the memory and for its length. 
The principle here can be explained as follows. To determine the state
of the link at time $t$, i.e. its presence (sampling of $X_t$ gives 1)
or is absence ($X_t$ gives 0), first we decide, 
with probability $q$, whether to copy one of the previous link states,
or to determine the presence of the link through a Bernoulli trial 
with probability $y$. When we draw a state from the past,
this state can be chosen in any way, but as we shall see in
the following, here for simplicity we pick uniformly from
the last $p$ steps of the time series.
In terms of random variables
this can be written as:
\begin{equation}
	X_t = Q_t X_{(t-Z_t )} + (1-Q_t)Y_t.
    \label{DARp_eqn}
\end{equation}
where, for each $t$, $Q_t \sim Bernoulli(q)$, $Y_t \sim Bernoulli(y)$ and $Z_t$ is
some random variable which picks integers in the range $(1,...,p)$. As
mentioned this $Z_t$ could take any form, indeed a natural choice may
be $Z_t \sim \exp(t)$, but for the sake of simplicity we here take
$Z_t \sim Uniform(1,p)$.  When $q=0$ the link has no memory, while for
$q \neq 0$ the process in Eq.~(\ref{DARp_eqn}) is clearly
non-Markovian, however, since the memory is finite, in that we only
consider $p$ previous values, we can view a DAR($p$) process as a
$p$-th order Markov chain with an enlarged space of states
\cite{Singer_PLOSONE14}.  We can then define the so-called
``$p$-state'' of link $(i,j)$ at time $t$, by combining the state of
the link at time $t$ along with its previous $p-1$ states as the
vector $\underline{S}_t^{ij}= \left(
X^{ij}_t,X^{ij}_{t-1},...,X^{ij}_{t-p+1}\right)$. The set of
$p$-states for each of the links is sufficient to completely describe
the dynamics of the network.  In a network with $N$ nodes generated by
our model, one can show that the expected degree $\left< k \right>$ of
a randomly chosen node at any point in time is given purely as a
function of $y$ as $\left<k\right> = y(N-1)$ (see Appendix A).
In summary, our model for temporal
networks, which we name Discrete Autoregressive Network   
model of order $p$, or DARN(p) model, 
depends on three parameters, namely: $y$, $q$ and $p$. 
The first parameter $y$ controls the density of the network. The second,
$q$, tunes the strength of the memory term in the process with respect
to the memoryless term. The final parameter, $p$, controls the length
of the memory, which can be thought of as the number of time steps
before the autocorrelation function decays exponentially (see
Appendix B for a discussion of the autocorrelation function, and
Appendix C for the initialisation of the network).

\begin{figure*}
\subfloat[\label{fig:full_dynamics}]{%
  \includegraphics[width = 0.5\textwidth]{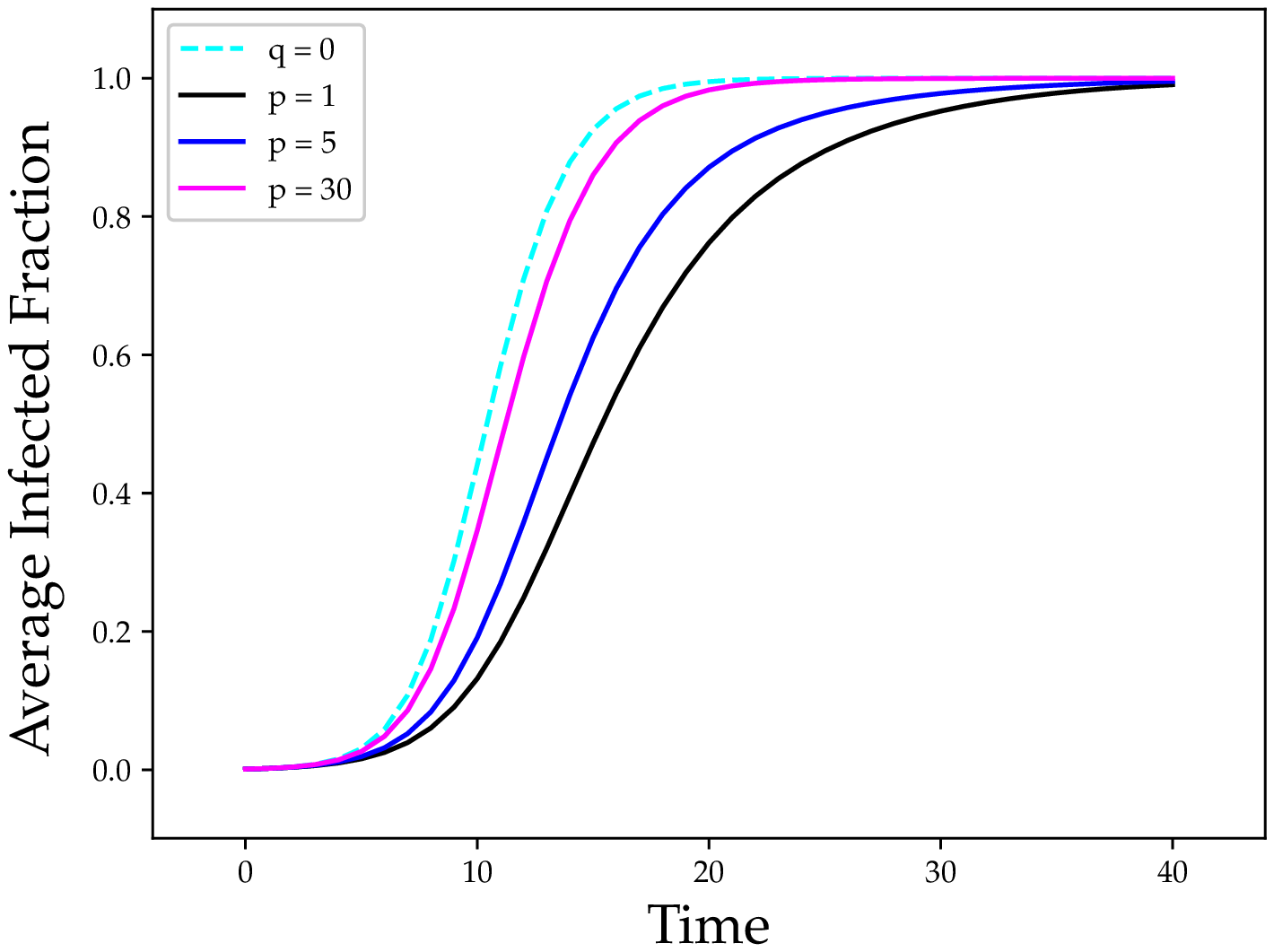}%
}\hfill
\subfloat[\label{fig:tte_values}]{%
  \includegraphics[width = 0.5\textwidth]{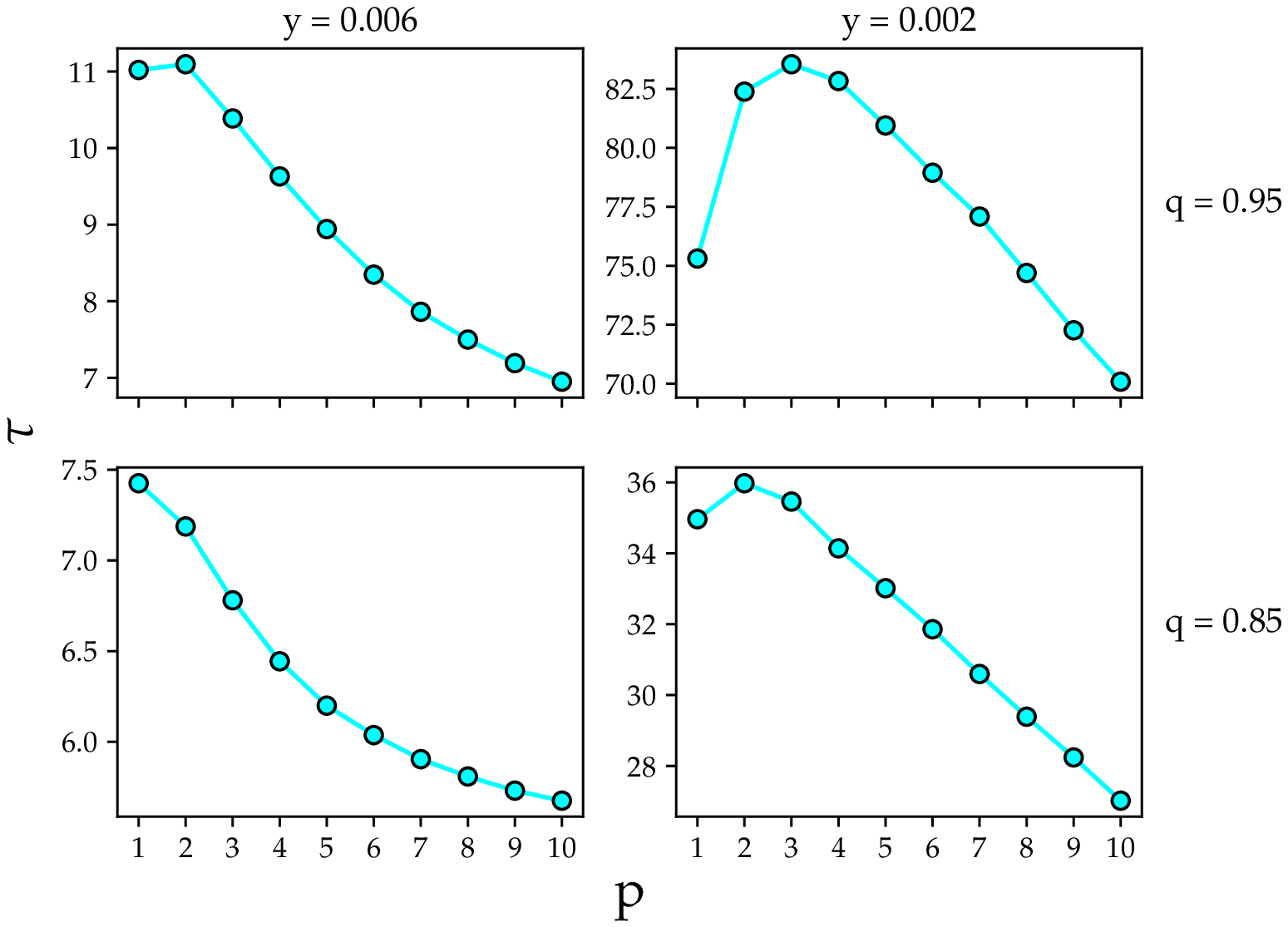}%
}
\caption{{\bf Effects of the memory length of a temporal network on a
    spreading process. a}, The average fraction of infected nodes over
  time for a disease spreading according to the SI model with $\lambda
  = 0.5$ on temporal networks with memory, as generated by the DARN(p) model
  in Eq.~(\ref{DARp_eqn}) with $y = 0.002$, $q=0.9$ and various
  values of $p$.  Each network has $N=1000$ nodes. Results are
  averaged over $10000$ realizations of the process.  {\bf b}, The
  average time $\tau$ until all the nodes in a temporal network are
  infected, given an SI model with $\lambda = 0.5$, is shown as a
  function of the memory length $p$ for $y=0.002$ and $y=0.006$, and
  $q =0.95$ and $q=0.85$.
    Each temporal network has $N = 1000$ nodes. Results are averaged over 100000 realization of the process. Resulting error bars are smaller than the markers. }
\label{}
\end{figure*}

\section{SPREADING PROCESSES ON TEMPORAL NETWORKS}
%
We have considered the simplest possible mechanism for propagating a
disease, or some message: the SI model, which is a special case of the SIS
model with a recovery rate of zero
\cite{Brauer2008,Prakash_ECMPPKDD10,kiss_book17}. 
We have adapted this for temporal networks by only allowing 
infection to pass between nodes at times when a link is present (see Appendix D).
Let us define ${\cal I}_t$ to be the set of infected nodes at a time
$t$. In our simulations we start with $\left| {\cal I}_0 \right| = 1$ and 
we study the dynamics of the epidemics for different values of the 
three parameters controlling our network model, namely the link density 
$y$, the strength of the memory $q$, and its length 
$p$.
In Fig. \ref{fig:full_dynamics} we plot the fraction ${\left| {\cal
    I}_{t} \right|}/{N}$ of infected nodes at time $t$ obtained, for
infectivity $\lambda = 0.5$, in a temporal network with $N=1000$ nodes produced
by a DARN(p) model with $y = 0.002$ and $q = 0.9$. This is compared to
the case of a temporal network without memory, generated by setting
$q=0$ in our model. We can see that the infection spreading in the
temporal network with no memory is faster than when any memory is
taken into account (i.e. when $q \neq 0$). In the cases where memory
is present, the infection spreading appears to depend heavily on both
$q$ and $p$. It is apparent that increasing the memory length $p$
changes how long it takes for the infection to spread across the
network, and that increasing $q$ exacerbates this behaviour. We also
observe that for large values of $p$ the curves converge to that of
the $q=0$ case, and this is in agreement with the fact that the
dynamics of our model in the limit of large $p$ are the same as those
of a random temporal network with no memory (see Appendix E).

In order to gain further insights into the spreading behaviour as a
function of the memory variables we quantify the speed of the
spreading by looking at the expected time taken until all of the nodes
in the network are infected. As before, we consider a temporal network
with $N=1000$ nodes and we fix an infection spreading rate $\lambda =
0.5$.  The SI model is run until the first time where all $N$ nodes
are infected. This value is then averaged over 100000 iterations of
the process to return the average time $\tau$ until full infection of
the network.  Fig. \ref{fig:tte_values} shows $\tau$ as a function of
the memory length $p$, and for different values of
$q$. For $y = 0.006$, we observe a monotonic decrease of $\tau$ as
  a function of $p$ when $q = 0.85$. However, when
  we increase the strength of the memory to $q = 0.95$,  
the time taken until the entire
  network is infected shows a non-monotonic dependence of the 
memory length, with the presence of  a maximum at $p_{\rm max}=2$.
  When we further decrease the network density to $y=0.002$, these maxima  
  move to $p_{\rm max} = 2$ and $p_{\rm max} = 3$
  respectively.

\section{THEORETICAL RESULTS}
%
Our model is in essence a generalisation of the ER random graph model
to the case of temporal networks with memory. As such, in our networks
there are no correlations between pairs of different links. Since each
link is independent, we can analyse the model by first analysing a
single link. The dynamics of a link can be explained in terms of the
transition matrix of the higher order Markov chain corresponding to
the DAR($p$) process. While the transition matrix for a first order
Markov chain expresses the probabilities of moving between the
possible states over a time step, namely here $(0 \to 0), (0 \to 1),
(1 \to 0)$ and $(1 \to 1)$, a $p$-th order transition matrix expresses
the probability of moving between $p$-states representing the possible
histories of the system.  If $\underline{S}_t= (X_t, X_{t-1}, \ldots
X_{t-p+1})$ is the $p$-state of our process $X_t$ at time $t$, then,
for any $\alpha, \beta \in \mathcal{S}$, where set $\mathcal{S}$ is
the set of all $2^p$ possible $p$-states, we can look at the
probability ${\rm Prob}(\underline{S}_{t+1} = \beta | \underline{S}_t
= \alpha)$. This defines the entries of the $p$-th order
$2^p\times2^p$ transition matrix $T_{\alpha \beta}$.
To write down this transition matrix it is useful to introduce an ordering
into the possible states of the system. Since there are $2^p$
possible states, we assign to each $\alpha \in \mathcal{S}$ a unique label
$l(\alpha) \in \left[ 0, 2^p -1 \right]$. To do this we note that by
 definition $\alpha = (\alpha_1,...,\alpha_p)$ with each $\alpha_i \in
 \{ 0,1 \}$. Hence, a convenient unique labelling is to take
 $\l(\alpha) = \sum_{i=1}^p \alpha_i 2^{p-i}$. For ease of notation,
 unless explicitly stated, $\alpha$ will refer to its label $l(\alpha)$. In
 this way we can
 write the $p$-th order transition matrix
 as:
\begin{eqnarray} \label{transition_mat}
	T_{\alpha \beta} = \left[q \frac{h(\alpha)}{p} +(1-q)y \right] \delta \left(\beta,2^{p-1} + \floor{\frac{\alpha}{2}}\right) +\nonumber \\
	 \left[1 - q \frac{h(\alpha)}{p} - (1-q)y \right] \delta \left(\beta,\floor{\frac{\alpha}{2}}\right), 
\end{eqnarray} 
where $h(x)$ is the Hamming weight of the number $x$ (the number of 1s
in its binary representation), $\delta(x,y) = 1$ if $x = y$ and $0$
otherwise, and $\floor{x}$ is the largest integer value smaller than
$x$.
Note also that, for the sake of simplicity, this matrix is
  indexed from zero, not one.
Taking two nodes, one of which is
initially infected, we study the expected time taken for the second
node to become infected. Since the infection process is modelled as a
Bernoulli random variable $\Lambda_t \sim Bernoulli(\lambda)$, the
infection is passed at the first value of $t$ such that $\Lambda_t X_t
= 1$. We can cast this process as a Markov chain by considering a
``dual-state" $(\underline{S}_t,\Lambda_t)$, where $\underline{S}_t$
and $\Lambda_t$ are the $p$-state of the link and its infectivity
state at time $t$, respectively. Let us call the set of all possible
dual-states $\tilde{\mathcal{S}}$, then we note that $\left| \tilde{\mathcal{S}}
\right| = 2^{p+1}$, and we set $(\alpha, \iota) = \tilde{\alpha}
\in \tilde{\mathcal{S}}$, where $\alpha$ is defined as for the transition matrix in Eq.~\ref{transition_mat} and $\iota = 1$ if an infection is passed and zero otherwise. The corresponding label function is then $\tilde{l}(\tilde{\alpha}) = l(\alpha) +
2^{p}\iota$. We can then define the transition matrix $P_{\tilde{\alpha}
 \tilde{\beta}}$ in block form as:
\begin{equation}
	\underline{\underline{P}} = 
	\begin{pmatrix}
		\left(1-\lambda \right) \underline{\underline{T}}	&	\lambda \underline{\underline{T}}\\
		\left(1-\lambda \right) \underline{\underline{T}}	&	\lambda \underline{\underline{T}}
	\end{pmatrix},
	\label{P_mat}
\end{equation}
where the sub-matrix $ \underline{\underline{T}}$ has elements given by Eq.~(\ref{transition_mat}).

\begin{figure*}
\subfloat[\label{sfig:link_theo_a}]{%
  \includegraphics[width = 0.5\textwidth]{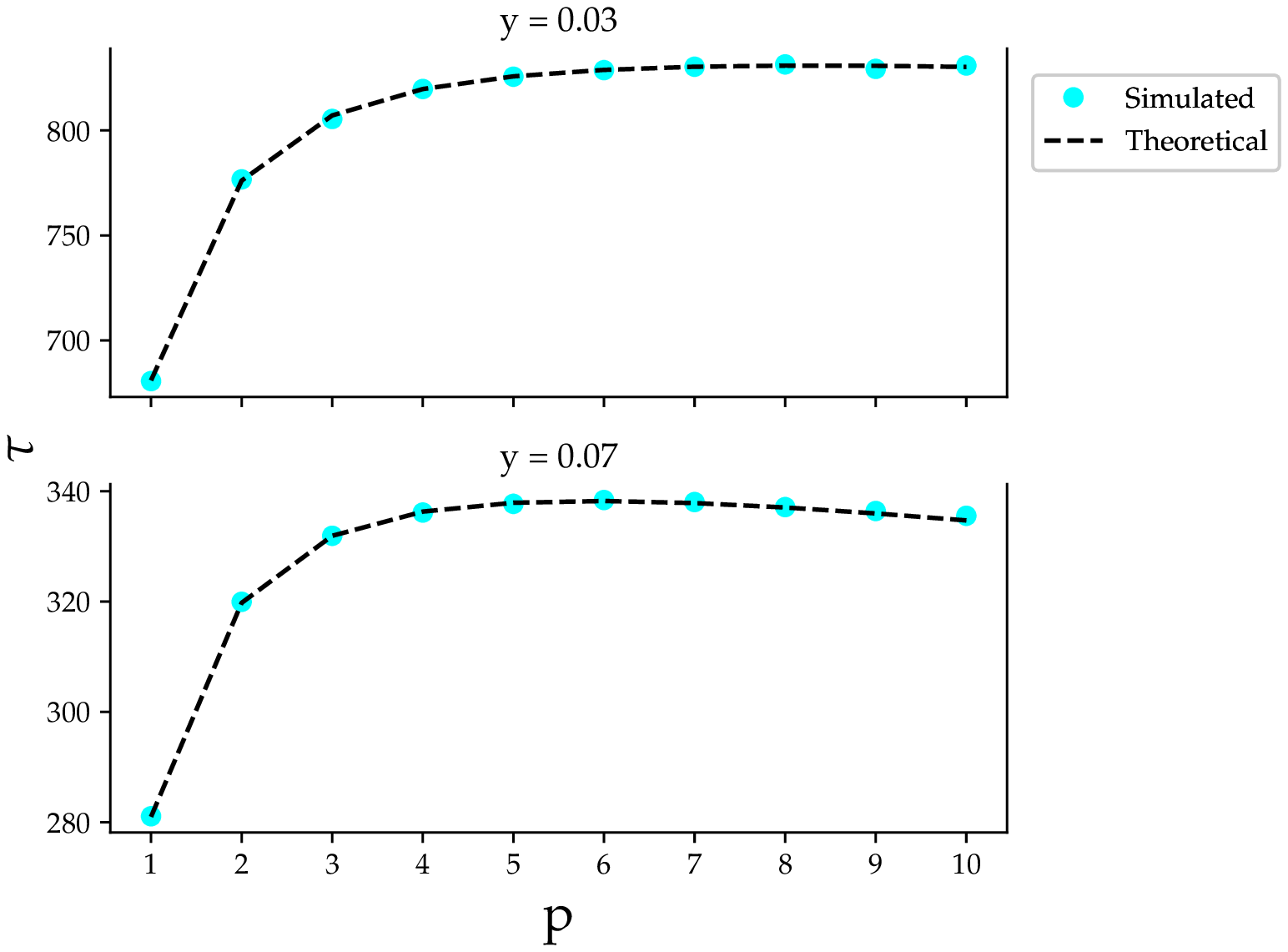}%
}\hfill
\subfloat[\label{sfig:link_theo_b}]{%
  \includegraphics[width = 0.5\textwidth]{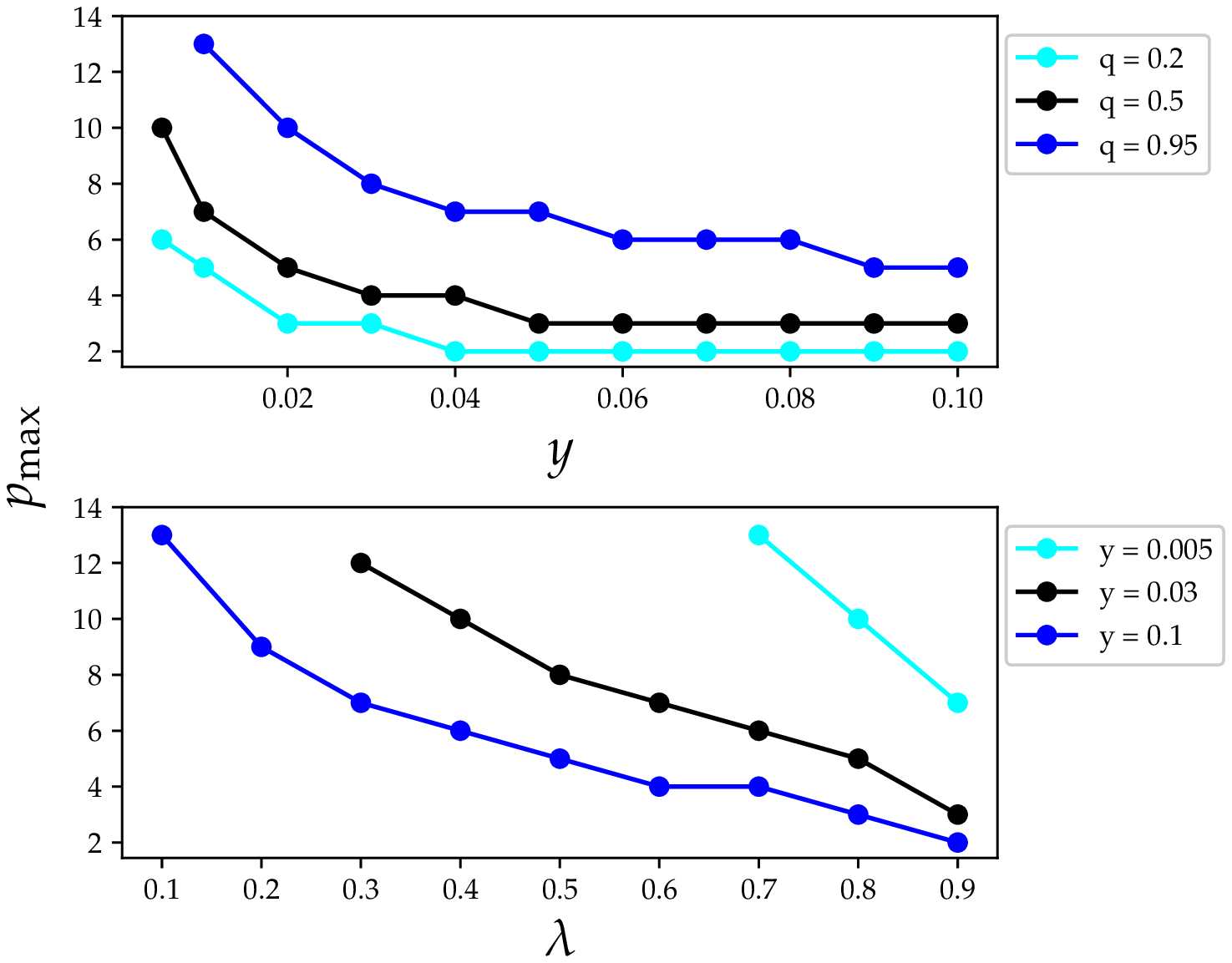}%
}
\caption{{\bf Theoretical results 
for the spreading over a single link. a,} The theoretical prediction of Eq.~(\ref{tau_avg_eqn}) is in good agreement with the simulated values of the average 
passage time $\tau$ of the infection over a single link 
as a function of the memory length $p$. 
Shown are cases with infectivity $\lambda = 0.5$, $q = 0.95$ and two values 
of $y$, namely $y = 0.03$ and $y = 0.07$. The average passage time $\tau$ as a function of $p$ has a maximum at the value $p = p_{\rm max}$, where $p_{\rm max}$ is dependent on $y$.
{\bf b,} The theoretical values for $p_{\rm max}$ are reported as a function of $y$ for infectivity $\lambda=0.5$ and for multiple values of $q$ (upper panel), and as a function of $\lambda$ for multiple values of $y$, when $q$ is fixed to 0.95 (lower panel).}
\label{}
\end{figure*}

Consider the set $A$ of dual-states where an infection is passed, i.e.
$A = \left\{ \tilde{\alpha} \in  \tilde{\mathcal{S}} : \tilde{\alpha}_1 = 1, \tilde{\alpha}_{p+1} = 1 \right\}$.
We are now interested in how long it takes our system
to reach a state in $A$. We can find the average hitting time
$\tau_{\tilde{\alpha}}$ from each starting state $\tilde{\alpha} \notin A$ as
the minimal solution of the following equations \cite{cinlar2013introduction}:
\begin{equation}
 \tau_{\tilde{\alpha}}  = 1 + \sum_{\tilde{\beta} \notin A}
 P_{\tilde{\alpha} \tilde{\beta}} \tau_{\tilde{\beta}}  .
	\label{avg_eqn}
\end{equation}
This is effectively saying that, if we start in state
$\tilde{\alpha}$, the average time taken to reach a state in $A$ is
the average time taken to reach a state in $A$ from any
$\tilde{\beta}$ weighted by the probability of moving to
$\tilde{\beta}$ from $\tilde{\alpha}$, plus the one time step it would
take to make that move. This equation can be
  simplified (see Appendix F), then we can average over the initial
  states of the system to get:
\begin{equation}
	\left< \tau  \right>_p = \sum_{\alpha=0}^{2^p-1} \tau _\alpha \text{ Prob}\left( l \left (\underline{S}_0 \right) = \alpha \right) ,
	\label{tau_avg_eqn}
\end{equation}
where the last term refers to the probability that the label of the initial $p$-state $\underline{S}_0$ of the process $X_t$ corresponds to the label of the $p$-state $\alpha$.

Eqs.~(\ref{avg_eqn}) and (\ref{tau_avg_eqn}) can be solved directly
for small values of $p$ (seeAppendix I for details). The plots in the
two panels of Fig. \ref{sfig:link_theo_a} show not only that the
theoretical predictions are in very good agreement with the
simulations, but also that the 
non-monotonic
behaviour observed in Fig. \ref{fig:tte_values} for networks with
$N=1000$ nodes can emerge even in the case of a single link.
In particular, we find that $\left< \tau \right>_p$ has a maximum at
$p_{\rm max} = 8$ when $y = 0.03$ and at $p_{\rm max} = 6$ when
$y = 0.07$.  Eq.~(\ref{tau_avg_eqn}) can be used to explore 
how $p_{\max}$ depends on the values of the parameters $y$ and $q$, and on the infectivity $\lambda$. 
In Fig.~(\ref{sfig:link_theo_b}) (upper panel) we see that, at fixed $\lambda$, the value of $p_{\rm max}$ decreases with $y$, while it increases with 
the memory strength $q$. In this way, for small $y$ and large $q$, the length of the memory $p$ which produces the 
maximal value of $\left< \tau \right>_p$ 
can be very large.  
For instance, when $y=0.01$ and $q=0.95$ we get $p_{\max}=13$. 
Fig.~(\ref{sfig:link_theo_b}) (lower panel) shows that $p_{\max}$ decreases with the infectivity of 
the SI process for each value of $y$. 

\section{PHASE DIAGRAM OF THE MODEL}
\begin{figure*}
\subfloat[\label{sfig:inflection_phase_1}]{%
  \includegraphics[width = 0.5\textwidth]{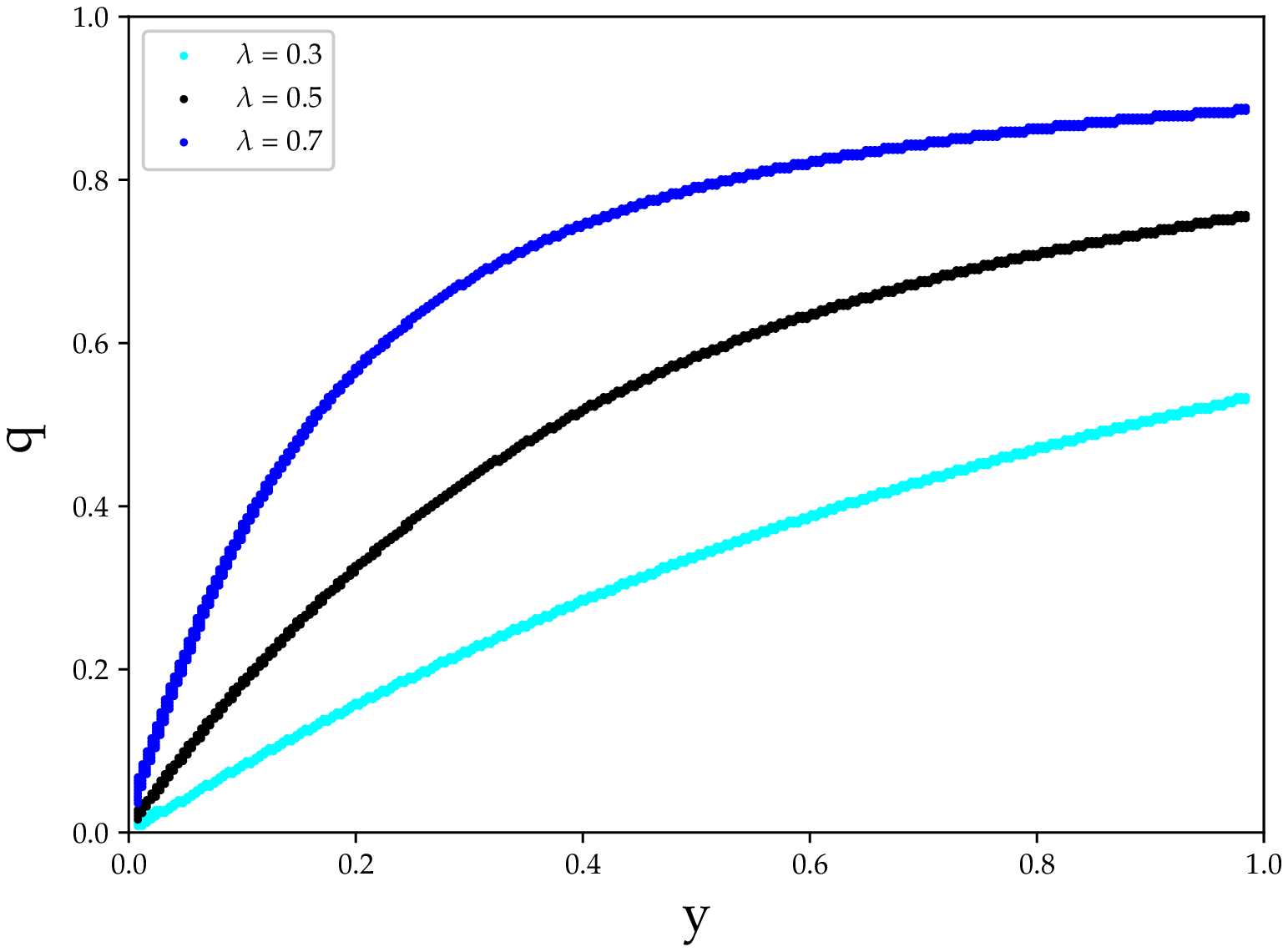}%
}\hfill
\subfloat[\label{sfig:inflection_phase_3}]{%
  \includegraphics[width = 0.5\textwidth]{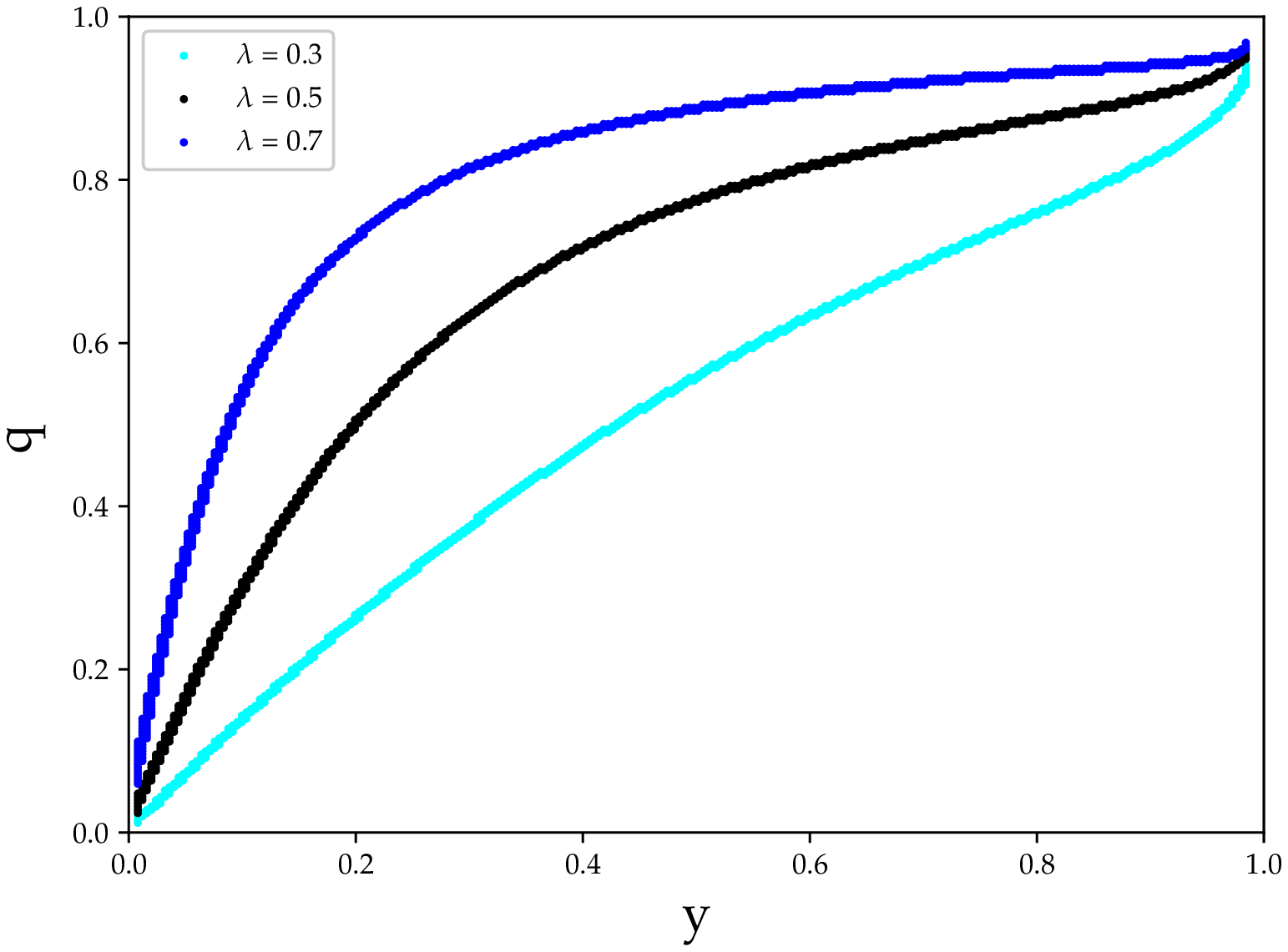}%
}
\caption{{\bf Phase diagram 
    for the presence of a non monotonic dependence of spreading time on memory 
    length}. For each value of the infectivity $\lambda$, the region above the curve denotes the values of the parameters $q,y$ where there must be a value of $p > 1$ such that $\tau$ has a local maximum. The left panel refers to the case of a single link, while the right  panel refers to the case of a DARN(p) model 
with $N=3$ nodes.
}
\label{fig:inflection}
\end{figure*}

%
We have found a solution to the average time taken for an infection to
pass across a single link, and hence along any chain of links. 
Real world networks, however, often contain many paths between any two
nodes, and these paths are often of varying lengths.
Eq.~(\ref{avg_eqn}) can be extended to deal with multiple paths of any length
(see Appendix G).  To do this, let us index each link in the network as
$\ell$, with $\ell=1,2,...,N(N-1)/2$. We can then define the dual-state of
the $\ell{\rm th}$ link as $\tilde{\alpha}^{\ell}$, as we did in the
single link case. The dual-state
of the entire network can then be written as
$\underline{\tilde{\alpha}} = (\tilde{\alpha}^1,\tilde{\alpha}^{2},...)$.
We can then generalise Eq.~(\ref{avg_eqn}) in
terms of $P_{\underline{\tilde{\alpha}} \underline{\tilde{\beta}}}$, which we refer
to as the transition tensor, between any two network dual-states, as
 \begin{equation}
	 \tau_{\underline{\tilde{\alpha}}}  = 1 + \sum_{\underline{\tilde{\beta}} \notin \mathcal{A}}
 	P_{\underline{\tilde{\alpha}} \underline{\tilde{\beta}}} \tau_{\underline{\tilde{\beta}}}.
	\label{avg_eqn_graph}
 \end{equation}
where $\mathcal{A}$ is the now the set
of network dual-states where we stop our infection process. 
In principle, this equation can be used for networks with any number
of nodes $N$, to study $p_{\rm max}$ as a function of $y$, $q$, and
$\lambda$, as was done in Fig.~\ref{sfig:link_theo_a}, although
computational constraints do not allow this for large $N$.  The
equation can be simplified to find the average passage time of an
infection over multiple paths of the same length (see Appendix H).
However, since we are interested in the most general case of multiple
paths with different lengths, let us now focus on a network with $N=3$
nodes, the smallest possible example of this type, having paths of
length 1 and length 2 between any two nodes. 
Eqs.~(\ref{avg_eqn_graph}) 
can be used to determine when a 
maximum in the time taken to pass an infection between any two nodes in a temporal network with $N=3$
should occur as a function of $p$. 
For each value of $\lambda$, by comparing the mean passage time for
$p=1,~2$ and $\infty$, we find sufficient conditions for the existence
of a maximum of $\left< \tau \right>_p$ at $p$ other than 1 (see
Appendix J). This defines a curve which breaks the $(q,y)$ plane into two
sections, the upper section being the one where $\left< \tau
\right>_p$ is non-monotonic.
These curves are displayed for $\lambda = 0.3,0.5$ and $0.7$ in
Fig. \ref{fig:inflection} for a single link (left) and for the $N=3$ node
network (right). The regions where a maximum must be present are
clearly dependent on $\lambda$ and on the number of nodes in the
network.  For $N=3$ we observe that approximately half of the
$(q,y)$ plane must result in a maximum for $\lambda = 0.3$,
however increasing $\lambda$ reduces this fraction. For example,
at a fixed value $y = 0.5$, when $\lambda = 0.3$
then nearly half of the possible $q$ values must produce a maximum,
when $\lambda = 0.5$ this fraction decreases to approximately 0.2, and
for $\lambda = 0.7$ then only 0.1 of the values of $q$ must result in
a maximum according to our criterion. The size of the regions
where a maximum must be present in general decreases with the number of
nodes in the network. Together with the shape of
the curves in Fig. \ref{fig:inflection}, this explains why in large and sparse 
networks, such as those considered in Fig. \ref{fig:tte_values},
we observe a non-monotonic behaviour of $\left< \tau \right>$ vs $p$
only for high memory strength $q$ and low graph density $y$.

\section{CONCLUSIONS}

Memory plays an important role in many processes in physics. In our
networked world, interactions change in time.  Such temporal changes
must be taken into account when studying dynamical processes, be they
the spreading of epidemics \cite{kiss_book17}, the diffusion of ideas
\cite{iacopini18}, the movement of people or patterns in broader
social interactions \cite{Rosvall_natcomm14}. 
The model we have introduced is a simple way to
include memory in a temporal network and can be further extended in
many directions, for instance to include correlations among links, or
to allow for different links to have different memory characteristics.
The results we have obtained, and the methods developed in doing so,
pave the way for a radical change in how we consider the influence of
memory in networks, and highlights how unexpected the results can be
when we do.

\begin{acknowledgments}
We acknowledge Piero Mazzarisi for interesting discussions.
V. L. acknowledges support from the EPSRC project EP/N013492/1.
FL acknowledges partial support by the grant SNS13LILLB ”Systemic risk in financial markets across time scales” and by the European Communitys H2020 Program under the scheme INFRAIA-1-2014-2015: Research Infrastructures, grant agreement \#654024 SoBigData: Social Mining and Big Data Ecosystem (http://www.sobigdata.eu).\\
\end{acknowledgments}

\section*{Author contributions}
All the authors designed the research. O.W. performed calculations and generated figures. All the authors discussed intermediate and final results and wrote the paper.

\section*{Competing interests}
The authors declare no competing interests, financial or otherwise.\\

\section*{Correspondance}
Correspondence and requests for materials should be addressed to O.W or V.L.

\appendix

\section{THE AVERAGE DEGREE OF A DAR($p$) NETWORK}
The average degree of a network with $N$ nodes, where each link is an independent DAR($p$) process, is its self a random process, defined by: 
\begin{equation}
	\left<k \right>_t = \frac{2}{N} \sum_{i} \sum_{j > i} X^{ij}_t .
\end{equation}
where $X_t= \{ X^{ij}_t \}$ is the adjacency matrix of the network
at time $t$. The average of this over time, which we simply denote by $\left< k
\right>$ is then: 
\begin{eqnarray}
\begin{aligned}
	\left< k \right> =& \frac{2}{N} \sum_{i} \sum_{j > i} {\overline{X^{ij}_t}}   \nonumber \\
	=& \frac{2}{N} \sum_{i} \sum_{j > i} y \nonumber \\
	=& y(N-1).
\end{aligned}
\end{eqnarray}
where the second line comes from the time averaged quantity
$\overline{X^{ij}_t}$ being
equal to $y$ \cite{jacobs1978discrete}.

\section{AUTOCORRELATION FUNCTIONS OF DAR($p$) PROCESSES}
\begin{figure}
	\includegraphics[width = 0.45\textwidth]{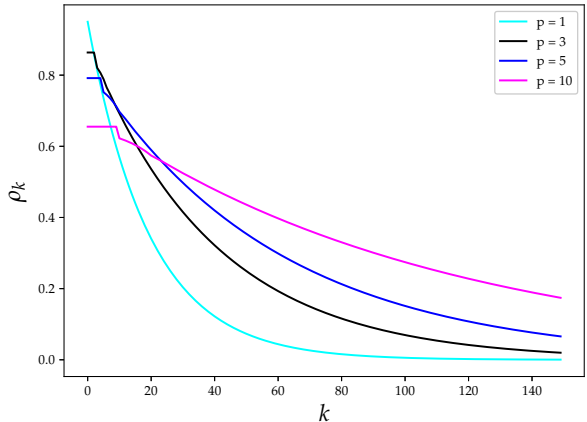}
	\caption{The autocorrelation function $\rho_k$ as a function of time
          $k$ for $q = 0.95$ and various values of $p$. The function is flat for the first $p$ time steps, and then decays exponentially.}
	\label{fig:autocorr}
\end{figure}

It is well known that DAR($p$) processes have an autocorrelation
function given by the Yule-Walker equations
\cite{jacobs1978discrete,macdonald1997hidden}.  The value $\rho_k$ of
the autocorrelation function at time shift $k$ is given by:
\begin{equation}
	\rho_k = \frac{q}{p} \sum^{p}_{i=1}  \rho_{k-i}.
	\label{Y_W_eqns}
\end{equation}
Using the facts that $\rho_0 = 1$ and $\rho_{-t} = \rho_{t}$ we find
the first $p$ values of $\rho_k$, i.e. the values for $k \leq
p$. Expanding and rearranging Eq~(\ref{Y_W_eqns}) gives
\begin{equation}
	\frac{p}{q} \rho_k - \sum^{k-1}_{i=1} \rho_i - \sum^{p-k}_{i=1} \rho_i = 1.
\end{equation}
We can then rewrite these equations in matrix form as
\begin{eqnarray}
\begin{aligned}
	\underline{\underline{M}} \, \underline{\rho} = & \underline{1} \\
	M_{ij} = & \frac{p}{q} \delta_{ij} - \delta(j \leq i-1) - \delta(j \leq p-i).
\end{aligned}
\end{eqnarray}
 We then notice the following:
\begin{equation}
\begin{aligned}
	\sum_j M_{ij} = & \frac{p}{q} - (i-1) - (p-i)=\\
	= & p \left( \frac{1}{q} -1 \right) + 1.
\end{aligned}
\end{equation}
Since this does not depend on $i$, we see that $\rho_k = \rho$ is a constant for all $k \leq p$, with: 
\begin{equation}
	\rho = \left( p \left( \frac{1}{q} -1 \right) + 1 \right)^{-1}.
\end{equation}
This then allows us to solve Eq.~(\ref{Y_W_eqns}) for the
autocorrelation function up to any finite time.
Resulting autocorrelation functions are shown in Fig.~\ref{fig:autocorr}
for various values of the parameters. 
\\

\section{INITIALISATION OF THE TEMPORAL NETWORK}

In our numerical simulations, we choose the initial condition of each
random variable by sampling a $p-$vector from the joint stationary
distribution of the DAR($p$) process.  This is done in the following
way We first sample each element of the $p$-state vector
$\underline{S}_0^{ij}$ of link $(i,j)$ at time $t=0$ from the
Bernoulli random variable $Y_t$.  We then allow the process in
Eq.~(\ref{DARp_eqn}) to run until the autocorrelation function of each
link has decayed below some threshold value. Once this point is
reached we consider this to be the state $\underline{S}_0^{ij}$ of
each link in the network.  This defines the time $t=0$ at which we
start to run any dynamical process on the network.  Any theoretical
results will use the steady state of the Markov chain directly.

\section{THE SI MODEL ON A TEMPORAL NETWORK}

In the SI model, the nodes of the network can be in one of two
possible states, namely susceptible (S) or infected (I). 
The infection can be passed over each link of the network
connecting an infected to a susceptible, with the infection process
modelled as a Bernoulli random variable $\Lambda_t \sim Bernoulli(\lambda)$.
We define
${\cal I}_t$ to be the set of infected nodes at a time $t$, and take
some initial subset ${\cal I}_0 \subset \cal{N}$ of nodes in the
infected state, then at each time $t$ for all infected nodes $i \in
{\cal I}_t$ each neighbouring node $j \in \partial i (t)$, where
$\partial j(t) \coloneqq \left\{ j \in {\cal N} : a^{ij}_{t} = 1
\right\}$ becomes infected with probability $\lambda$.  We take all
infection spreading to happen simultaneously on the network at time
$t$. If there are multiple infected neighbours to a susceptible node
then they will all attempt to pass an infection, and the susceptible
node will become infected if at least one of these neighbours
succeeds. Since there is no recovery, this change in state in the SI
model is permanent.

\section{THE LONG MEMORY LIMIT OF THE NETWORK}
To explore the limiting behaviour of our model, let us first consider the conditional 
probability 
\begin{multline}
	P \left( X_t = 1 | X_{t-1} = x_{t-1},...., X_{t-p} = x_{t-p} \right)  \\
	= (1-q)y + q \phi_t (p),
\end{multline}
where
\begin{equation}
	\phi_t (p) = \frac{1}{p} \sum_{i=1}^{p} x_{t-i}.
	\label{mem_kern}
\end{equation}
%
%

Since Eq.~(\ref{mem_kern}) is a sample expectation of a stationary
DAR($p$) process, and $\overline{X_t} = y$ \cite{jacobs1978discrete},
we can see that $\phi_t(p) \to y$ as $p \to \infty$.
Now it remains to prove that
fluctuations away from the stationary state can be ignored. Consider
the time series for $\phi_t (p)$, as given by
\begin{equation}
	\phi_{t+1} (p) = \phi_t (p) + \frac{x_t - x_{t-p}}{p}. 
\end{equation}
Since $x_t - x_{t-p} \in \{-1,0,1\}$ we can then set the following bounds:
\begin{equation}
	y - \frac{t}{p} \leq \phi_t (p) \leq y + \frac{t}{p}.
\end{equation}
Given that the passage of an infection over a link in our model must
happen in finite time (except in trivial cases such as $\lambda = 0$)
then we know that as $p \to \infty$ we must have that $\phi_t (p) \to
y$.  Given that the conditional probability of observing a link does
not change in time, the large $p$ limit of the DAR($p$) process is
then indistinguishable from a Bernoulli process with probability
$y$. Since each link in our network is independent then each link will
follow a Bernoulli process, and so the network becomes a memoryless
random network with link probability $y$.

\section{AVERAGE PASSAGE TIME FOR A SINGLE LINK}
Eq.~\ref{avg_eqn} can be solved directly as a set of $2^{p+1}$ linear
equations.  Given that this grows rapidly with $p$, we wish to make
use of some inherent symmetry to reduce the number of equations, and
thereby allow us to find solutions for higher values of p.

Consider Eq.~\ref{avg_eqn}, we know that for all $j \in A$ we have $j > 2^p$, so this can be broken up into
\begin{equation}
	\tau_i^A = 1 + \sum_{j \leq 2^p} P_{ij} \tau_j^A + \sum_{j > 2^p ; j \notin A} P_{ij} \tau_j^A .
\end{equation}
Looking at our matrix $P$ in Eq.~\ref{P_mat} we can re-write this as:
\begin{equation}
	\tau_i^A = 1 + \sum_{j \leq 2^p} (1 - \lambda) T_{ij} \tau_j^A + \sum_{j < 2^{p-1}} \lambda T_{ij} \tau_j^A .
\end{equation}
As before we note that all the elements of $A$ are memory states with leading value 1, so we can define the matrix $T^L_{ij} = T_{ij}$ if $j < 2^{p-1}$ and 0 else. This allows us to write
\begin{equation}
	\tau_i^A = 1 + \sum_{j \leq 2^p} (1 - \lambda) T_{ij} \tau_j^A + \sum_{j < 2^p} \lambda T^L_{ij} \tau_j^A .
\end{equation}
Hence we only need to solve for the first $2^p$ values, allowing us to write the equation in matrix form as:
\begin{equation}
	\underline{\tau} = \underline{1} + \left( (1-\lambda) \underline{\underline{T}} + \lambda \underline{\underline{T}}^L \right) \underline{\tau} .
\end{equation}
Giving
\begin{equation}
	\underline{\tau} = \left( \underline{\underline{Id}} - (1-\lambda) \underline{\underline{T}} - \lambda \underline{\underline{T}}^L \right)^{-1} \underline{1} .
\end{equation}
Then, defining the matrix $\underline{\underline{\Phi}}$ as: 
\begin{eqnarray}
\begin{aligned}
	\Phi_{\alpha \beta} = \left( 1- \lambda \right) &\left[q \frac{h(\alpha)}{p} +(1-q)y \right] \delta \left(\beta,2^{p-1} + \floor{\frac{\alpha}{2}}\right) \\
	 + &\left[1 - q \frac{h(\alpha)}{p} - (1-q)y \right] \delta \left(\beta,\floor{\frac{\alpha}{2}}\right). 
\end{aligned}
\end{eqnarray}
We can write
\begin{equation}
	\tilde{\underline{\tau}}  = \left( \underline{\underline{Id}} - \underline{\underline{\Phi}} \right)^{-1} \underline{1},
	\label{tau_eqn}
\end{equation}
This greatly simplifies any calculation of average passage times for single links.

\section{TEMPORAL NETWORKS DESCRIBED BY TENSORS}
In this work we model each link in a network as a Markov chain, and so each link has a 
transition matrix associated with it. If each Markov chain has a state space $\mathcal{S}$ 
with $\left| \mathcal{S} \right| = \sigma$ then the temporal network with $N$ nodes 
can be described as a Markov chain with state space $\mathcal{S}^N$ (meaning the 
cartesian product of $\mathcal{S}$ with its self $N$ times) with 
$\left| \mathcal{S}^N \right| = \sigma^N$. Rather than attempt to directly impose an 
ordering on the states and generate a transition matrix we instead form a transition 
tensor of rank $2N$. This tensor is formed by having one source index and one 
target index for each link in the network; each index represents a label for a state in 
$\mathcal{S}$, and so varies from 1 to $N$ (or 0 to $N-1$). We then define 
the transition tensor $T$ as 
\begin{equation}
	T_{\underline{\alpha} \underline{\beta}} = P \left(\bigcap_{i = 1,...,N} \alpha_i \to \beta_i \right).
\end{equation}
With $\underline{\alpha}$ and $\underline{\beta}$ the sets of source and target 
indices for each link.\\
We use this approach to transform quantities derived from the transition matrix 
into a tensor form in the same way. 
When calculating the average hitting time of a Markov chain, we do 
so from some starting state, as labeled by a single index. 
For our new formulation, each state is labeled by a vector of indices, 
and so each possible starting state is labeled by a vector of indices. 
In this way 
Eq.~(\ref{avg_eqn}) becomes
\begin{equation}
	\left< \tau \right>_{\underline{\alpha}} = 1 + \sum_{\beta \notin \mathcal{A}} T_{\underline{\alpha} \underline{\beta}} \left< \tau \right>_{\underline{\beta}},
	\label{tensor_avg}
\end{equation}
as in Eq.~(\ref{avg_eqn_graph}). The two equations can be seen to become equivalent 
upon flattening Eq.~(\ref{tensor_avg}) so that $T$ becomes an $\sigma^N \times \sigma^N$ matrix 
and $ \left< \tau \right>$ becomes a vector of length $\sigma^N$.

\section{AVERAGE PASSAGE TIME FOR MULTIPLE LINKS IN PARALLEL}

Solving for the average passage time over a single link allows us to
extrapolate the average passage time along any number of links in
series by using the linearity of expectation. Real systems however
will often have multiple paths, eventually of different lengths. Here
we examine the simplest case of $m$ paths of unit length in
parallel. Naively solving Eq.~(\ref{avg_eqn_graph}) for the case of
$m$ links in parallel requires us to handle a $2^{m(p+1)} \times
2^{m(p+1)}$ matrix. However, in the same way as with the single link,
we can greatly simplify this. We wish to find the average time taken
for an infection to pass across any of $m$ direct links from a source
node to a target node. Let us start with the tensor equation
Eq.~(\ref{avg_eqn_graph}). If we write $\underline{\tilde{\alpha}}_i=
(\alpha_i, \iota_i^s)$ and $\underline{\tilde{\beta}}_i= (\beta_i,
\iota_i^t)$ then
\begin{eqnarray}
\begin{aligned}
	 \tau_{\underline{\tilde{\alpha}}}  =& 1 + \sum_{\underline{\tilde{\beta}} \notin \mathcal{A}} \,P_{\underline{\tilde{\alpha}} \underline{\tilde{\beta}}} \tau_{\underline{\tilde{\beta}}}\\
	=& 1 + \sum_{(\underline{\beta},\underline{\lambda}) \notin \mathcal{A}}
	T_{\underline{\alpha}\underline{\beta}}\, \Lambda_{\underline{\iota}^s, \underline{\iota}^t}\, \tau_{\underline{\beta}}.
\end{aligned}
\end{eqnarray}
Here the tensors $T_{\underline{\alpha}\underline{\beta}}$ and $ \Lambda_{\underline{\iota}^s \underline{\iota}^t}$ 
are defined by
\begin{eqnarray}
\begin{aligned}
	T_{\underline{\alpha}\underline{\beta}} =& \prod_{i=1}^{m} T_{\alpha_i, \beta_i}\\
	 \Lambda_{\underline{\iota}^s \underline{\iota}^t} =& \lambda^{h(\underline{\iota}^t)} \left(1-\lambda \right)^{m - h(\underline{\iota}^t)}.
\end{aligned}
\end{eqnarray}
Where $h(\underline{\iota}^t)$ is the number of ones in $\underline{\iota}^t$. Let us now assert that 
no link will start in an infecting state, and so we $\tau_{(\underline{\alpha},\underline{\iota}^1)} = \tau_{(\underline{\alpha},\underline{\iota}^2)}$ for any two $\underline{\iota}^1$ and $\underline{\iota}^2$, hence we may write $ \tau_{\underline{\tilde{\alpha}}}$ as $ \tau_{\underline{\alpha}}$.Then we introduce two index partitioning sets $I_1$ and $I_1$ so that $I_1 \subset \{1,....,m\}$ and $\{1,....,m\} = I_1 \cup I_2$. So that
\begin{equation}
	 \tau_{\underline{\alpha}}  = 1 + \sum_{I_1,I_2} \sum_{\beta_i \in I_1} \sum_{\beta_i \in I_2} \sum_{\underline{\iota}^t} 
	T_{\underline{\alpha}\underline{\beta}}\, \Lambda_{\underline{\iota}^s \underline{\iota}^t}\, \tau_{\underline{\beta}}
	\chi\left( (\underline{\beta},\underline{\iota}^t) \notin \mathcal{A} \right).
\end{equation}  
With $\chi$ being the indicator function. Due to the lack of dependence of $\Lambda$ on its source state, 
we can write this as
\begin{equation}
	 \tau_{\underline{\alpha}}  = 1 + \sum_{I_1,I_2} \sum_{\beta_i \in I_1 \leq 2^{p-1} } \sum_{\beta_i \in I_2 > 2^{p-1}}
	 T_{\underline{\alpha}\underline{\beta}}\, \tau_{\underline{\beta}} \, L(I_1,I_2),
\end{equation}
for some function $L(I_1,I_2)$, by noticing that we only care about the value of $\iota_i^t$ if $\beta_i > 2^{(p-1)}$. 
It is then straightforward to show that $L(I_1,I_2) = (1-\lambda)^{\left| I_2 \right|}$. We can now take our definition 
of $\underline{\underline{T}}^L$ and define $\underline{\underline{T}}^R = \underline{\underline{T}} - \underline{\underline{T}}^L$ 
and write
\begin{eqnarray}
\begin{aligned}
	 \tau_{\underline{\alpha}}  =& 1 + \sum_{I_1,I_2} \sum_{\underline{\beta}} 
	 \left( \prod_{\beta_i \in I_1} T^L_{\alpha_i \beta_i} \right)  \left( \prod_{\beta_i \in I_2} T^R_{\alpha_i \beta_i} \right) 
	 (1-\lambda)^{\left| I_2 \right|} \tau_{\underline{\beta}}\\
	 =& 1 + \sum_{\underline{\beta}} \prod_{i} \left( T^L_{\alpha_i \beta_i} + (1-\lambda) T^R_{\alpha_i \beta_i} \right)\\
	 =& 1 + \sum_{\underline{\beta}} \prod_{i} \Phi_{\alpha_i \beta_i}.
\end{aligned}
\end{eqnarray}
Giving us an equation in terms of elements of a single $2^p \times 2^p$ matrix.

\section{AVERAGE PASSAGE TIME FOR A SMALL NETWORK}
We wish to find the average passage time for an infection between any two nodes in a three node complete network.
To do this, we must first write down 
$P_{\underline{\tilde{\alpha}} \underline{\tilde{\beta}}}$. First we define the set of possible dual-states $H$ 
of a link where an infection is passed. Then define the ``infected" transition matrix
$P'_{\tilde{\alpha} \tilde{\beta}}$ to be $P_{\tilde{\alpha} \tilde{\beta}}$ (as in Eqn. \ref{P_mat}) if $\tilde{\alpha} \notin H$, and $P'_{\tilde{\alpha} \tilde{\beta}} = \delta_{\tilde{\alpha} \tilde{\beta}}$ otherwise, and the "waiting" transition matrix $\tilde{P}_{\tilde{\alpha} \tilde{\beta}} = T_{\alpha \beta}$ if $\tilde{\alpha} = (\alpha,0)$ and $\tilde{\beta} =(\beta,0)$ and  $\tilde{P}_{\tilde{\alpha} \tilde{\beta}} = \delta_{\tilde{\beta} (\alpha,0)}$ if $\tilde{\alpha} = (\alpha,1)$. We then write 
\begin{equation}
\begin{aligned}
	P_{\underline{\tilde{\alpha}} \underline{\tilde{\beta}}} = &P'_{\tilde{\alpha}^1 \tilde{\beta}^1} \left(\chi_H(\tilde{\alpha}^1)P'_{\tilde{\alpha}^2 \tilde{\beta}^2} + (1-\chi_H(\tilde{\alpha}^1)) \tilde{P}_{\tilde{\alpha}^2 \tilde{\beta}^2}  \right) P'_{\tilde{\alpha}^3 \tilde{\beta}^3} \\
	\label{tau_3_eqn}
\end{aligned}
\end{equation}
Where links 1 and 3 are connected to the infection source, and links 2 and 3 are connected to the infection target. Our set $A$ from Eqn. \ref{avg_eqn} has now become $A = \{ \underline{\tilde{\alpha}} : \alpha^2, \alpha^3 \in H \}$.

\section{CONDITIONS FOR THE EXISTENCE OF MAXIMUM POINTS}

Equations (\ref{avg_eqn}) and 
(\ref{avg_eqn_graph}) for the average passage time 
can be solved numerically for $p = 1$ 
and $p = 2$. We can then use the obtained values, namely 
$\left< \tau  \right>_1$ and $\left< \tau  \right>_2$,    
along with the solution $\left< \tau  \right>_{\infty}$ in
the $p \to \infty$ limiting case, to look for values of $\lambda$, $q$ 
and $y$ that must result in a non-monotonic behaviour of 
$\left< \tau  \right>_p$ as a function of $p$. These occur when 
$\left< \tau  \right>_1 < \left< \tau  \right>_2$ and $\left< \tau  \right>_{\infty} < \left< \tau  \right>_2$.
It is important to note that this is only a sufficient condition for
the existence of inflection points. Whilst when this condition holds there must be 
non monotonic behaviour in $\left< \tau  \right>_{p}$, it is possible 
to observe non monotonicity without it, so what we can extract in this way
is only a subset of the possible cases where maxima will be observed. 


\end{document}